\documentclass[10pt,a4paper,english,twocolumn]{IEEEtran}
\usepackage[T1]{fontenc}
\usepackage[latin9]{inputenc}
\usepackage{amsmath}
\usepackage{amsthm}
\usepackage{amssymb}
\usepackage{stackrel}
\usepackage{graphicx}

\makeatletter

\pdfpageheight\paperheight
\pdfpagewidth\paperwidth

\theoremstyle{plain}
\newtheorem{thm}{\protect\theoremname}
\theoremstyle{plain}
\newtheorem{lem}[thm]{\protect\lemmaname}
\theoremstyle{remark}
\newtheorem{rem}[thm]{\protect\remarkname}

\usepackage{subfigure}
\usepackage{epstopdf}
\usepackage{cite}

\usepackage{balance}

\makeatother

\usepackage{babel}
\providecommand{\lemmaname}{Lemma}
\providecommand{\remarkname}{Remark}
\providecommand{\theoremname}{Theorem}

\begin{document}

\title{Performance Analysis of Dense Small Cell Networks with Generalized
Fading}

\author{\IEEEauthorblockN{Bin Yang\IEEEauthorrefmark{1}, Ming Ding\IEEEauthorrefmark{2}, Guoqiang
Mao\IEEEauthorrefmark{4}\IEEEauthorrefmark{2}\IEEEauthorrefmark{1},
Xiaohu Ge\IEEEauthorrefmark{1}}\\
\IEEEauthorblockA{\IEEEauthorrefmark{1}School of Electronic Information and Communications,
Huazhong University of Science and Technology, China.}\\
\IEEEauthorblockA{\IEEEauthorrefmark{2}Data61, CSIRO, Australia.}\\
\IEEEauthorblockA{\IEEEauthorrefmark{4}School of Computing and Communication, The University
of Technology Sydney, Australia.}\\
\IEEEauthorblockA{Email: \IEEEauthorrefmark{1}\{yangbin, xhge\}@hust.edu.cn, \IEEEauthorrefmark{2}ming.ding@data61.csiro.au,
\IEEEauthorrefmark{4}g.mao@ieee.org.}}
\maketitle
\begin{abstract}
In this paper, we propose a unified framework to analyze the performance
of dense small cell networks (SCNs) in terms of the coverage probability
and the area spectral efficiency (ASE). In our analysis, we consider
a practical path loss model that accounts for both non-line-of-sight
(NLOS) and line-of-sight (LOS) transmissions. Furthermore, we adopt
a generalized fading model, in which Rayleigh fading, Rician fading
and Nakagami-$m$ fading can be treated in a unified framework. The
analytical results of the coverage probability and the ASE are derived,
using a generalized stochastic geometry analysis. Different from existing
work that does not differentiate NLOS and LOS transmissions, our results
show that NLOS and LOS transmissions have a significant impact on
the coverage probability and the ASE performance, particularly when
the SCNs grow dense. Furthermore, our results establish for the first
time that the performance of the SCNs can be divided into four regimes,
according to the intensity (aka density) of BSs, where in each regime
the performance is dominated by different factors. 
\end{abstract}

\section{Introduction}

With increasing demands on wireless data driven by smartphones, tablets,
and video streaming, wireless data traffic is expected to overwhelm
cellular networks in the near future. Against this background, network
densification, together with millimeter wave and massive multiple-input
multiple-output, have been envisioned to be ``the three pillars''
to support the vision of the emerging fifth generation (5G) wireless
networks in the future to accommodate the phenomenal growth of wireless
data traffic \cite{LopezPerez15Towards}. In this context, the orthogonal
deployment of dense small cellular networks (SCNs) \cite{Ge165G,Ge15Energy}
has been selected as the workhorse for capacity enhancement in the
fourth generation (4G) and the 5G networks developed by the third
Generation Partnership Project (3GPP). In this paper, we focus on
the analysis of such dense SCNs.

Different from most previous work studying network performance where
the propagation path loss between the base staions (BSs) and the mobile
users (MUs) follows the same power-law model, in this paper we consider
the co-existence of both non-line-of-sight (NLOS) and line-of-sight
(LOS) transmissions, which frequently occur in \emph{urban areas}.
More specifically, for a randomly selected MU, BSs deployed according
to a homogeneous Poisson point process (PPP) are divided into two
categories, i.e., NLOS BSs and LOS BSs, depending on the distance
between the BSs and the MU. In this context, the authors of \cite{Ding16Performance}
studied the coverage and capacity performance by using a multi-slop
path loss model incorporating probabilistic NLOS and LOS transmissions.
The authors of \cite{DiRenzo15StochasticJ} and \cite{Singh15Tractable}
analyzed the coverage and capacity performance in millimeter wave
cellular networks. In \cite{DiRenzo15StochasticJ}, a three-state
statistical model for each link was assumed, in which a link can either
be in a NLOS, LOS or in an outage state. In \cite{Singh15Tractable},
self-backhauled millimeter wave cellular networks were characterized
assuming a cell association scheme based on the smallest path loss.
However, both \cite{DiRenzo15StochasticJ} and \cite{Singh15Tractable}
assumed a noise-limited network, ignoring inter-cell interference,
which may not be very practical since modern wireless networks generally
work in an interference-limited region. 

In this work, we will study the performance impact on the coverage
probability and the ASE caused by the coexistence of both NLOS and
LOS transmissions in dense SCNs with a generalized fading model. Furthermore,
the proposed framework can also be applied to analyze dense SCNs,
where BSs are distributed according to non-homogeneous PPPs, i.e.,
the BS intensity (aka density) varies in the spatial domain. The main
contributions of this paper are as follows:
\begin{itemize}
\item A general SCN model: For characterizing the signal-to-interference-plus-noise
ratio (SINR) coverage probabiliy and the ASE in SCNs, a unified framework
is proposed, which is applicable to analyze a SCN assuming a user
association scheme based on the strongest received signal power, with
a generalized fading channel model and considering both NLOS and LOS
transmissions.
\item Theoretical analysis: The coverage probability and the ASE are derived
based on the proposed model incorporating LOS/NLOS transmissions and
generalized fading. The accuracy of our analytical results is validated
by Monte Carlo simulations.
\item Performance insights: Different from existing work that does not differentiate
NLOS and LOS transmissions, our analysis reveals distinctly different
results that both SINR and SIR distributions depend on BS intensity.
Furthermore, our results establish for the first time that the performance
of SCNs can be divided into four regimes, according to the intensity
of BSs, where in each regime the performance is dominated by different
factor. 
\end{itemize}
The reminder of this paper is organized as follows. Section \ref{sec:System-Model}
introduces the system model and network assumptions. Section \ref{sec:The-Coverage-Probability}
studies the coverage probability and the ASE in a SCN. In Section
\ref{sec:Results-and-Discussions}, the analytical results are validated
via Monte Carlo simulations and the performance of SCNs with different
fading models is investigated. Finally, Section \ref{sec:Conclusions-and-Future}
concludes this paper and discusses possible future work.

\section{\label{sec:System-Model}System Model}

We consider a homogeneous SCN in urban areas and focus on the analysis
of downlink performance. We assume that BSs are spatially distributed
on the infinite plane and the locations of BSs $\boldsymbol{X}_{i}$
follow independent homogeneous Poisson point processes (PPPs) denoted
by $\Phi=\left\{ \boldsymbol{X}_{i}\right\} $ with an intensity of
$\lambda$, where $i$ is the BS index. MUs are deployed according
to another homogeneous point process denoted by $\Phi_{u}$ with an
intensity of $\lambda_{u}$. All BSs in the network operate at the
same power $P_{t}$ and share the same bandwidth. Within a cell, MUs
use orthogonal frequencies for downlink access and therefore \emph{intra-cell
interference} is not considered in our analysis. However, adjacent
BSs may generate \emph{inter-cell interference} to MUs, which is the
main focus of our work. 

\subsection{\label{subsec:Signal-Propagation-Model}Path Loss Model}

We incorporate both NLOS and LOS transmissions into the path loss
model, which recently attracts growing attentions among researches.
In reality, the occurrence of NLOS or LOS transmissions depends on
various environmental factors, including geographical structure, distance
and clusters, etc. The following definition gives a simplified one-parameter
model of NLOS and LOS transmissions.

The occurrence of NLOS and LOS transmissions can be modeled by probabilities
$p^{\textrm{NL}}\left(R_{i}\right)$ and $p^{\textrm{L}}\left(R_{i}\right)$,
respectively. They are functions of the distance between a BS $\boldsymbol{X}_{i}$
and the typical MU, which satisfy
\begin{equation}
p^{\textrm{NL}}\left(R_{i}\right)+p^{\textrm{L}}\left(R_{i}\right)=1,
\end{equation}
where $R_{i}=\left\Vert \boldsymbol{X}_{i}\right\Vert $ denotes the
Euclidean distance between a BS at $\boldsymbol{X}_{i}$ and the typical
MU (alternatively called the probe MU or the tagged MU) located at
the origin $o$.

Regarding the mathematical form of $p^{\textrm{L}}\left(R_{i}\right)$
(or $p^{\textrm{NL}}\left(R_{i}\right)$), N. Blaunstein \cite{Blaunstein98Parametric}
formulated $p^{\textrm{L}}\left(R_{i}\right)$ as a negative exponential
function, i.e., $p^{\textrm{L}}\left(R_{i}\right)=e^{-\kappa R_{i}}$,
where $\kappa$ is a parameter determined by the intensity and the
mean length of the blockages lying in the visual path between the
typical MU and the serving BS. Bai \cite{Bai14Analysis} extended
N. Blaunstein's work by using the random shape theory which shows
that $\kappa$ is not only determined by the mean length but also
the mean width of the blockages. The authors of \cite{Ding16Performance}
considered $p^{\textrm{L}}\left(R_{i}\right)$ as a linear function
and a two-piece exponential function, respectively, which are both
recommended by the 3GPP.

It should be noted that the NLOS (or LOS) probability is assumed to
be independent for different links from BSs to the typical MU. Though
such assumption might not be entirely realistic in some scenarios
as links may be spatially correlated, the authors of \cite{Bai14Analysis}
showed that it causes negligible loss of accuracy in the SINR analysis.

Note that from the viewpoint of the typical MU, each BS on the infinite
plane $\mathbb{R}^{2}$ is either a NLOS BS or a LOS BS. Accordingly,
we perform a thinning procedure on points in the PPP $\Phi$ to model
the distributions of NLOS BSs and LOS BSs, respectively. That is,
each BS in $\Phi$ will be kept if a BS has a NLOS transmission with
the typical MU, thus forming a new point process denoted by $\Phi^{\textrm{NL}}$.
While BSs in $\Phi\setminus\Phi^{\textrm{NL}}$ form another point
process denoted by $\Phi^{\textrm{L}}$, representing the set of BSs
with LOS path to the typical MU. As a consequence of the independence
assumption of LOS and NLOS transmissions mentioned in the last paragraph,
$\Phi^{\textrm{NL}}$ and $\Phi^{\textrm{L}}$ are two independent
non-homogeneous PPPs with intensity functions $\lambda p^{\textrm{NL}}\left(R_{i}\right)$
and $\lambda p^{\textrm{L}}\left(R_{i}\right)$, respectively.

In general, NLOS and LOS transmissions incur different path losses
and different fadings, which are captured by the following equations
\begin{equation}
P_{i}^{\textrm{NL}}=P_{t}A^{\textrm{NL}}h^{\textrm{NL}}\left(R_{i}\right)^{-\alpha^{\textrm{NL}}}=B^{\textrm{NL}}h^{\textrm{NL}}\left(R_{i}\right)^{-\alpha^{\textrm{NL}}}\label{eq:Power_N}
\end{equation}
and
\begin{equation}
P_{i}^{\textrm{L}}=P_{t}A^{\textrm{L}}h^{\textrm{L}}\left(R_{i}\right)^{-\alpha^{\textrm{L}}}=B^{\textrm{L}}h^{\textrm{L}}\left(R_{i}\right)^{-\alpha^{\textrm{L}}}\label{eq:Power_L}
\end{equation}
where $A^{\textrm{NL}}$ and $A^{\textrm{L}}$ are the path losses
at a reference distance for NLOS and LOS transmissions, respectively,
$B^{\textrm{NL}}=P_{t}A^{\textrm{NL}}$ and $B^{\textrm{L}}=P_{t}A^{\textrm{L}}$
are both constants, $h^{\textrm{NL}}$ and $h^{\textrm{L}}$ are random
variables capturing the channel power gains for NLOS and LOS transmissons
from the BS $\boldsymbol{X}_{i}$ to the typical MU, respectively.
Therefore, the received power of the typical MU from BS $\boldsymbol{X}_{i}$
is given by
\begin{equation}
P_{i}\left(R_{i}\right)=\mathbb{I}_{i}P_{i}^{\textrm{NL}}+\left(1-\mathbb{I}_{i}\right)P_{i}^{\textrm{L}},\label{eq:rec_power}
\end{equation}
where $\mathbb{I}_{i}$ is a random indicator variable, which equals
to 1 for a NLOS transmission and 0 for a LOS transmission, and the
corresponding probabilities are $p^{\textrm{NL}}\left(R_{i}\right)$
and $p^{\textrm{L}}\left(R_{i}\right)$, respectively, i.e.,
\begin{equation}
\mathbb{I}_{i}=\begin{cases}
1, & \hspace{-0.3cm}\textrm{with probability }p^{\textrm{NL}}\left(R_{i}\right)\\
0, & \hspace{-0.3cm}\textrm{with probability }p^{\textrm{L}}\left(R_{i}\right)
\end{cases}.\label{eq:indicator}
\end{equation}

Based on the path loss model discussed above, for downlink transmissions,
the SINR experienced by the typical MU associated with BS $\boldsymbol{X}_{i}$
can be written by
\begin{align}
\textrm{SINR}_{i} & =\frac{S}{I+\eta}=\frac{P_{i}\left(R_{i}\right)}{\underset{\boldsymbol{X}_{z}\in\Phi\setminus\boldsymbol{X}_{i}}{\sum}P_{z}\left(R_{z}\right)+\eta},\label{eq:SINR}
\end{align}
where $\Phi\setminus\boldsymbol{X}_{i}$ is the Palm point process
\cite{Baccelli09Stochastic} representing the set of interfering BSs
in the network to the typical MU and $\eta$ denotes the noise power
at the MU side, which is assumed to be additive white Gaussian noise
(AWGN).

\subsection{Cell Association Scheme}

Considering NLOS and LOS transmissions, the typical MU should connect
with the BS that provides the highest SINR. Such BS does not necessarily
have to be the the nearest BS from the typical MU in the SCN. More
specifically, the typical MU associates itself to the BS $\boldsymbol{X}_{i}^{*}$
given by
\begin{align}
\boldsymbol{X}_{i}^{*} & =\arg\underset{\boldsymbol{X}_{i}\in\Phi}{\max}\textrm{SINR}_{i}.\label{eq:maxSINR}
\end{align}
Intuitively, the highest SINR association is equivalent to the strongest
received signal power association. Such intuition is formally presented
and proved in Lemma \ref{lem: lem1}.
\begin{lem}
\label{lem: lem1}For a non-negative set $\Xi=\left\{ a_{q}\right\} $,
$q\in\mathbb{N}$, then $\frac{a_{m}}{\underset{q\neq m}{\sum}a_{q}+W}>\frac{a_{n}}{\underset{q\neq n}{\sum}a_{q}+W}$
if and only if $a_{m}>a_{n}$, $\forall a_{m},a_{n}\in\Xi$.
\end{lem}
\begin{IEEEproof}
For a non-negative set $\Xi=\left\{ a_{q}\right\} $, $q\in\mathbb{N}$,
$\frac{a_{m}}{\underset{q}{\sum}a_{q}+W}>\frac{a_{n}}{\underset{q}{\sum}a_{q}+W}$
if and only if $a_{m}>a_{n}$, thus $\frac{a_{m}}{\underset{q}{\sum}a_{q}+W-a_{m}}>\frac{a_{n}}{\underset{q}{\sum}a_{q}+W-a_{n}}$,
which completes the proof.
\end{IEEEproof}
Lemma \ref{lem: lem1} states that providing the highest SINR is equivalent
to providing the strongest received power to the typical MU. Using
a similar mathematical form as Eq. \eqref{eq:maxSINR}, the typical
MU associates itself to the BS $\left(\boldsymbol{X}_{i},\textrm{U}\right)^{*}$
given by
\begin{align}
\left(\boldsymbol{X}_{i},\textrm{U}\right)^{*} & =\arg\underset{(\boldsymbol{X}_{i},\textrm{U})\in\mathbb{S}}{\max}B^{\textrm{U}}\mathcal{H}^{\textrm{U}}\left(R_{i}\right)^{-\alpha^{\textrm{U}}},\label{eq:maxrecpower}
\end{align}
where $\boldsymbol{X}_{i}\in\Phi$, $\textrm{U}\in\left\{ \textrm{NL},\textrm{L}\right\} $
and the set $\mathbb{S}=\Phi\times\left\{ \textrm{NL},\textrm{L}\right\} $.
In the following, we mainly use Eq. \eqref{eq:maxrecpower} to characterize
the considered cell association scheme.

\section{\label{sec:The-Coverage-Probability}The Coverage Probability and
the ASE Analysis}

In downlink performance evaluation, theoretical studies of the typical
MU located at the origin $o$ is sufficient to characterize the performance
of cellular networks where BSs are random distributed according to
a PPP \cite{Baccelli09Stochastic}. In this section, the coverage
probability is firstly investigated and then the ASE will be derived
from the results of coverage probability.

\subsection{General Case and Main Result}

In general, the coverage probability is defined as the probability
that the typical MU's measured SINR is greater than a designated threshold
$T$, i.e.,
\begin{equation}
p_{c}\left(\lambda,T\right)=\Pr\left[\textrm{SINR}>T\right],
\end{equation}
where the definition of SINR is given by Eq. \eqref{eq:SINR} and
the subscript $i$ is omitted here for simplicity. Now, we present
the main theorem on the coverage probability assuming a generalized
fading model as follows.
\begin{thm}[Coverage Probability]
\label{thm:Pcoverage } If we denote $\overline{R_{i}^{\textrm{NL}}}=R_{i}\cdot\left(B^{\textrm{NL}}h^{\textrm{NL}}\right)^{-1/\alpha^{\textrm{NL}}}$
and $\overline{R_{i}^{\textrm{L}}}=R_{i}\cdot\left(B^{\textrm{L}}h^{\textrm{L}}\right)^{-1/\alpha^{\textrm{L}}}$,
$\left\{ \overline{R_{i}^{\textrm{NL}}}\right\} _{i\in\mathbb{N}}$
(or $\left\{ \overline{R_{i}^{\textrm{L}}}\right\} _{i\in\mathbb{N}}$)
remains a PPP denoted by $\overline{\Phi^{\textrm{NL}}}$ (or $\overline{\Phi^{\textrm{L}}}$
) according to the displacement theorem \cite{Baccelli09Stochastic}.
Given that the signal propagation model follows Eq. \eqref{eq:rec_power}
and the typical MU selects the serving BS according to Eq. \eqref{eq:maxrecpower},
the SINR coverage probability $p_{c}\left(\lambda,T\right)$ assuming
a generalized fading can be evaluated by
\begin{equation}
p_{c}\left(\lambda,T\right)=p_{c}^{\textrm{NL}}\left(\lambda,T\right)+p_{c}^{\textrm{L}}\left(\lambda,T\right),\label{eq:theorem_pc}
\end{equation}
where
\begin{align}
 & p_{c}^{\textrm{NL}}\left(\lambda,T\right)=\int_{y=0}^{\infty}\int_{\omega=-\infty}^{\infty}\left[\frac{1-e^{-j\omega/T}}{2\pi j\omega}\right]\lambda^{\textrm{NL}}\left(y\right)\nonumber \\
 & \quad\,\times\exp\biggl\{-\Lambda^{\textrm{L}}\left(\left[0,y^{\alpha^{\textrm{NL}}/\alpha^{\textrm{L}}}\right]\right)-\Lambda^{\textrm{NL}}\left(\left[0,y\right]\right)+j\omega\eta y^{\alpha^{\textrm{NL}}}\nonumber \\
 & \quad\,+\int_{t=y^{\alpha^{\textrm{NL}}/\alpha^{\textrm{L}}}}^{\infty}\left[e^{j\omega y^{\alpha^{\textrm{NL}}}t^{-\alpha^{\textrm{L}}}}-1\right]\lambda^{\textrm{L}}\left(t\right)\textrm{d}t\nonumber \\
 & \quad\,+\int_{t=y}^{\infty}\left[e^{j\omega\left(y/t\right)^{\alpha^{\textrm{NL}}}}-1\right]\lambda^{\textrm{NL}}\left(t\right)\textrm{d}t\biggr\}\textrm{d}\omega\textrm{d}y,\label{eq:theorem_pcN}
\end{align}
and
\begin{align}
 & p_{c}^{\textrm{L}}\left(\lambda,T\right)=\int_{y=0}^{\infty}\int_{\omega=-\infty}^{\infty}\left[\frac{1-e^{-j\omega/T}}{2\pi j\omega}\right]\lambda^{\textrm{L}}\left(y\right)\nonumber \\
 & \quad\,\times\exp\biggl\{-\Lambda^{\textrm{NL}}\left(\left[0,y^{\alpha^{\textrm{L}}/\alpha^{\textrm{NL}}}\right]\right)-\Lambda^{\textrm{L}}\left(\left[0,y\right]\right)+j\omega\eta y^{\alpha^{\textrm{L}}}\nonumber \\
 & \quad\,+\int_{t=y^{\alpha^{\textrm{L}}/\alpha^{\textrm{NL}}}}^{\infty}\left[e^{j\omega y^{\alpha^{\textrm{L}}}t^{-\alpha^{\textrm{NL}}}}-1\right]\lambda^{\textrm{NL}}\left(t\right)\textrm{d}t\nonumber \\
 & \quad\,+\int_{t=y}^{\infty}\left[e^{j\omega\left(y/t\right)^{\alpha^{\textrm{L}}}}-1\right]\lambda^{\textrm{L}}\left(t\right)\textrm{d}t\biggr\}\textrm{d}\omega\textrm{d}y\label{eq:theorem_pcL}
\end{align}
where the intensity measures and intensities of $\overline{\Phi^{\textrm{NL}}}$
and $\overline{\Phi^{\textrm{L}}}$ are
\begin{equation}
\Lambda^{\textrm{NL}}\left(\left[0,t\right]\right)=\mathbb{E}_{h^{\textrm{NL}}}\left[2\pi\lambda\int_{R_{i}=0}^{t\left(B^{\textrm{NL}}h^{\textrm{NL}}\right)^{1/\alpha^{\textrm{NL}}}}p^{\textrm{NL}}\left(R_{i}\right)R_{i}\textrm{d}R_{i}\right],\label{eq:Measure_N}
\end{equation}
\begin{equation}
\Lambda^{\textrm{L}}\left(\left[0,t\right]\right)=\mathbb{E}_{h^{\textrm{L}}}\left[2\pi\lambda\int_{R_{i}=0}^{t\left(B^{\textrm{L}}h^{\textrm{L}}\right)^{1/\alpha^{\textrm{L}}}}p^{\textrm{L}}\left(R_{i}\right)R_{i}\textrm{d}R_{i}\right],\label{eq:Measure_L}
\end{equation}
\begin{equation}
\lambda^{\textrm{NL}}\left(t\right)=\frac{\textrm{d}}{\textrm{d}t}\Lambda^{\textrm{NL}}\left(\left[0,t\right]\right),\label{eq:Intensity_N}
\end{equation}
and
\begin{equation}
\lambda^{\textrm{L}}\left(t\right)=\frac{\textrm{d}}{\textrm{d}t}\Lambda^{\textrm{L}}\left(\left[0,t\right]\right),\label{eq:Intensity_L}
\end{equation}
where $j=\sqrt{-1}$ denotes the imaginary unit.
\end{thm}
\begin{IEEEproof}
See Appendix.
\end{IEEEproof}
Note that Theorem \ref{thm:Pcoverage } applies to a general case
with generalized fading. We now turn our attention to a few relevant
special cases where NLOS transmissions and LOS transmissions are concatenated
with Nakagami-$m$ fading of different parameters, which are more
practical in the real SCNs.

\subsection{Special Case 1: NLOS and LOS Transmissions are concatenated with
Nakagami-$m$ Fading}

In this subsection, we assume that both NLOS and LOS transmissions
(or signal amplitudes) are concatenated with Nakagami-$m$ fadings
of different parameters, e.g., $m^{\textrm{NL}}$ and $m^{\textrm{L}}$,
then the channel power gains are distributed according to Gamma distributions.
That is, 
\begin{equation}
f_{H^{\textrm{U}}}\left(h\right)=\frac{\left(m^{\textrm{U}}\right)^{m^{\textrm{U}}}}{\Gamma\left(m^{\textrm{U}}\right)}h^{m^{\textrm{U}}-1}e^{-m^{\textrm{U}}h},
\end{equation}
where $\textrm{U}\in\left\{ \textrm{NL},\textrm{L}\right\} $. Moreover,
a simplified NLOS/LOS transmission model is used for a specific analysis,
which is expressed by
\begin{equation}
p^{\textrm{L}}\left(R_{i}\right)=\begin{cases}
1, & \hspace{-0.3cm}R_{i}\in\left(0,d\right]\\
0, & \hspace{-0.3cm}R_{i}\in\left(d,\infty\right]
\end{cases},
\end{equation}
where $d$ is a constant distance below which all BSs connect with
the typical MU with LOS transmissions. This model has been commonly
used for modeling NLOS/LOS transmissions \cite{DiRenzo15StochasticJ,Bai15Coverage,Singh15Tractable}.

By substituting the PDF of $h^{\textrm{NL}}$ and $h^{\textrm{L}}$
into Eq. \eqref{eq:Measure_N} \textendash{} Eq. \eqref{eq:Intensity_L},
the intensity measures and intensities of $\overline{\Phi^{\textrm{NL}}}$
and $\overline{\Phi^{\textrm{L}}}$ can be readily obtained as follows

\begin{align}
 & \Lambda^{\textrm{NL}}\left(\left[0,t\right]\right)=-\frac{\pi\lambda d^{2}}{\Gamma\left(m^{\textrm{NL}}\right)}\Gamma\left(m^{\textrm{NL}},\frac{m^{\textrm{NL}}}{B^{\textrm{NL}}}\left(\frac{d}{t}\right)^{\alpha^{\textrm{NL}}}\right)\nonumber \\
 & +\frac{\pi\lambda t^{2}}{\Gamma\left(m^{\textrm{NL}}\right)}\left(\frac{B^{\textrm{NL}}}{m^{\textrm{NL}}}\right)^{\frac{2}{\alpha^{\textrm{NL}}}}\Gamma\left(\frac{2}{\alpha^{\textrm{NL}}}+m^{\textrm{NL}},\frac{m^{\textrm{NL}}}{B^{\textrm{NL}}}\left(\frac{d}{t}\right)^{\alpha^{\textrm{NL}}}\right),\label{eq:Measure_N_Nakagami}
\end{align}
\begin{align}
 & \Lambda^{\textrm{L}}\left(\left[0,t\right]\right)=\frac{\pi\lambda d^{2}}{\Gamma\left(m^{\textrm{L}}\right)}\Gamma\left(m^{\textrm{L}},\frac{m^{\textrm{L}}}{B^{\textrm{L}}}\left(\frac{d}{t}\right)^{\alpha^{\textrm{L}}}\right)\nonumber \\
 & +\frac{\pi\lambda t^{2}}{\Gamma\left(m^{\textrm{L}}\right)}\left(\frac{B^{\textrm{L}}}{m^{\textrm{L}}}\right)^{\frac{2}{\alpha^{\textrm{L}}}}\gamma\left(\frac{2}{\alpha^{\textrm{L}}}+m^{\textrm{L}},\frac{m^{\textrm{L}}}{B^{\textrm{L}}}\left(\frac{d}{t}\right)^{\alpha^{\textrm{L}}}\right),\label{eq:Measure_L_Nakagami}
\end{align}
\begin{align}
\lambda^{\textrm{NL}}\left(t\right) & =\frac{2\pi\lambda t}{\Gamma\left(m^{\textrm{NL}}\right)}\left(\frac{B^{\textrm{NL}}}{m^{\textrm{NL}}}\right)^{\frac{2}{\alpha^{\textrm{NL}}}}\nonumber \\
 & \quad\,\times\Gamma\left(\frac{2}{\alpha^{\textrm{NL}}}+m^{\textrm{NL}},\frac{m^{\textrm{NL}}}{B^{\textrm{NL}}}\left(\frac{d}{t}\right)^{\alpha^{\textrm{NL}}}\right),\label{eq:Intensity_N_Nakagami}
\end{align}
and
\begin{equation}
\lambda^{\textrm{L}}\left(t\right)=\frac{2\pi\lambda t}{\Gamma\left(m^{\textrm{L}}\right)}\left(\frac{B^{\textrm{L}}}{m^{\textrm{L}}}\right)^{\frac{2}{\alpha^{\textrm{L}}}}\gamma\left(\frac{2}{\alpha^{\textrm{L}}}+m^{\textrm{L}},\frac{m^{\textrm{L}}}{B^{\textrm{L}}}\left(\frac{d}{t}\right)^{\alpha^{\textrm{L}}}\right),\label{eq:Intensity_L_Nakagami}
\end{equation}
respectively, where $\Gamma\left(s,x\right)=\int_{x}^{\infty}v^{s-1}e^{-v}\textrm{d}v$
and $\gamma\left(s,x\right)=\int_{0}^{x}v^{s-1}e^{-v}\textrm{d}v$
denote the upper and the lower incomplete gamma functions, respectively,
$\Gamma\left(s\right)=\int_{0}^{\infty}v^{s-1}e^{-v}\textrm{d}v$
is the gamma function. The intermediate steps are easy to derive and
thus omitted here. By incorporating Eq. \eqref{eq:Measure_N_Nakagami}
- \eqref{eq:Intensity_L_Nakagami} into Eq. \eqref{eq:theorem_pcN}
and Eq. \eqref{eq:theorem_pcL}, the coverage probability of a SCN
experiencing Nakagami-$m$ fading can be calculated.

\subsection{Special Case 2: NLOS Transmission + Rayleigh Fading and LOS Transmission
+ Rician Fading}

In this part, we consider a more common case in which NLOS transmission
and LOS transmission are concatenated with Rayleigh fading and Rician
fading, respectively, i.e., $h^{\textrm{NL}}$ follows an exponential
distribution and $h^{\textrm{L}}$ follows a noncentral Chi-squared
distribution. With $m=\left(K+1\right)^{2}/2K+1$, Rician fading can
be approximated by a Nakagami-$m$ distribution, where $K$ is the
Rician $K$-factor representing the ratio between the power of the
direct path and that of the scattered paths. Without loss of generality,
we assume $f_{H^{\textrm{NL}}}\left(h\right)=e^{-h}$ and $f_{H^{\textrm{L}}}\left(h\right)=\frac{m^{m}}{\Gamma\left(m\right)}h^{m-1}e^{-mh}$
for NLOS and LOS transmissions, respectively.

As we have provided the intensity measure and intensity of $\overline{\Phi^{\textrm{L}}}$
experiencing Nakagami-$m$ fading in the previous subsection, in this
part we just provide the intensity measures and intensities of $\overline{\Phi^{\textrm{NL}}}$.
By substituting the PDF of $h^{\textrm{NL}}$ into Eq. \eqref{eq:Measure_N}
and Eq. \eqref{eq:Intensity_N}, $\Lambda^{\textrm{NL}}\left(\left[0,t\right]\right)$
and $\lambda^{\textrm{NL}}\left(t\right)$ can be easily evaluated
by
\begin{align}
\Lambda^{\textrm{NL}}\left(\left[0,t\right]\right) & =\pi\lambda t^{2}\left(B^{\textrm{NL}}\right)^{\frac{2}{\alpha^{\textrm{NL}}}}\Gamma\left(\frac{2}{\alpha^{\textrm{NL}}}+1,\frac{1}{B^{\textrm{NL}}}\left(\frac{d}{t}\right)^{\alpha^{\textrm{NL}}}\right)\nonumber \\
 & -\pi\lambda d^{2}\exp\left[-\frac{\left(d/t\right)^{\alpha^{\textrm{NL}}}}{B^{\textrm{NL}}}\right],
\end{align}
and
\begin{equation}
\lambda^{\textrm{NL}}\left(t\right)=2\pi\lambda t\left(B^{\textrm{NL}}\right)^{\frac{2}{\alpha^{\textrm{NL}}}}\Gamma\left(\frac{2}{\alpha^{\textrm{NL}}}+1,\frac{1}{B^{\textrm{NL}}}\left(\frac{d}{t}\right)^{\alpha^{\textrm{NL}}}\right),
\end{equation}
respectively. After substituting the intensity measures and intensities
of $\overline{\Phi^{\textrm{NL}}}$ and $\overline{\Phi^{\textrm{L}}}$
into Eq. \eqref{eq:theorem_pcN} and Eq. \eqref{eq:theorem_pcL},
the coverage probability can be obtained and we omit the rest derivations.

\subsection{The ASE Analysis}

Finally, we turn our attention to the ASE in the unit of $\textrm{bps/Hz/k\ensuremath{m^{2}}}$
for a given BS intensity $\lambda$, which can be evaluated as follows
\begin{align}
\textrm{ASE}\left(\lambda\right) & =\lambda\cdot\mathbb{E}_{\textrm{SINR}}\left[\log_{2}\left(1+\textrm{SINR}\right)\right]\nonumber \\
 & =\frac{\lambda}{\ln2}\int_{u=0}^{\infty}\frac{p_{c}\left(\lambda,u\right)}{u+1}\textrm{d}u.\nonumber \\
 & \overset{\left(a\right)}{\approx}\frac{\lambda}{\ln2}\stackrel[n=1]{N_{G}}{\sum}\frac{w_{n}}{u_{n}+1}p_{c}\left(\lambda,u_{n}\right),\label{eq:ASE}
\end{align}
where the approximation in $(a)$ follows the Gauss-Chebyshev Quadrature
(GCQ) rule \cite[Eq. (10)]{Yilmaz12AUnified} which makes numerical
computation easier, $u_{n}$ and $w_{n}$ are defined by
\begin{equation}
u_{n}=\tan\left[\frac{\pi}{4}\cos\left(\frac{2n-1}{2N_{G}}\pi\right)+\frac{\pi}{4}\right],
\end{equation}
and
\begin{equation}
w_{n}=\frac{\pi^{2}\sin\left(\frac{2n-1}{2N_{G}}\pi\right)}{4N_{G}\cos^{2}\left[\frac{\pi}{4}\cos\left(\frac{2n-1}{2N_{G}}\pi\right)+\frac{\pi}{4}\right]},
\end{equation}
respectively, where the truncation index $N_{G}$ should be set to
a sufficiently large value, such as 30, to ensure a good accuracy
\cite{Yilmaz12AUnified} .

\section{\label{sec:Results-and-Discussions}Results and Discussions}

In this section, we present numerical results to validate the accuracy
of our theoretical analysis, followed by discussions to shed new light
on the performance of the SCNs. We use the following parameter values,
$P_{t}=24\textrm{ dBm}$, $A^{\textrm{NL}}=10^{-3.29}$, $A^{\textrm{L}}=10^{-4.14}$,
$\alpha^{\textrm{NL}}=3.75$, $\alpha^{\textrm{L}}=2.09$, $\eta=-95\textrm{ dBm}$
and $d=250\textrm{ m}$ \cite{Singh15Tractable,3GPP36828,Kannan08Robust,Yang16Coverage}. 

\subsection{Validation of the Analytical Results of $p_{c}\left(\lambda,T\right)$
with Monte Carlo Simulations}

\begin{figure}
\begin{centering}
\includegraphics[width=8.3cm]{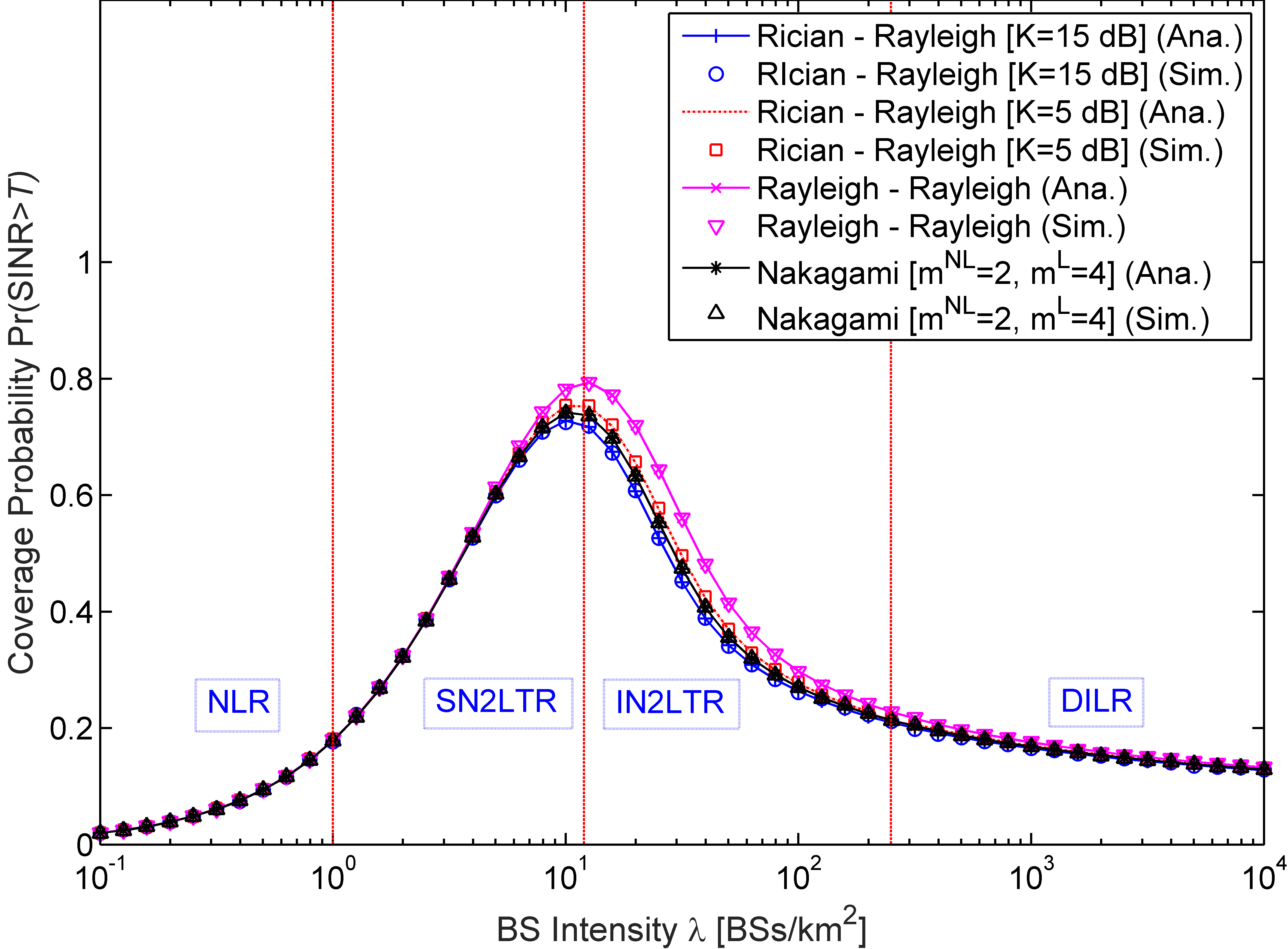}
\par\end{centering}
\caption{\label{fig:SINR-Coverage-Probability-density1}Coverage probability
vs. BS intensity $\lambda$, $\eta=-95\textrm{ dBm}$, $T=0\textrm{ dB}.$}
\end{figure}

\begin{figure}
\begin{centering}
\includegraphics[width=8.3cm]{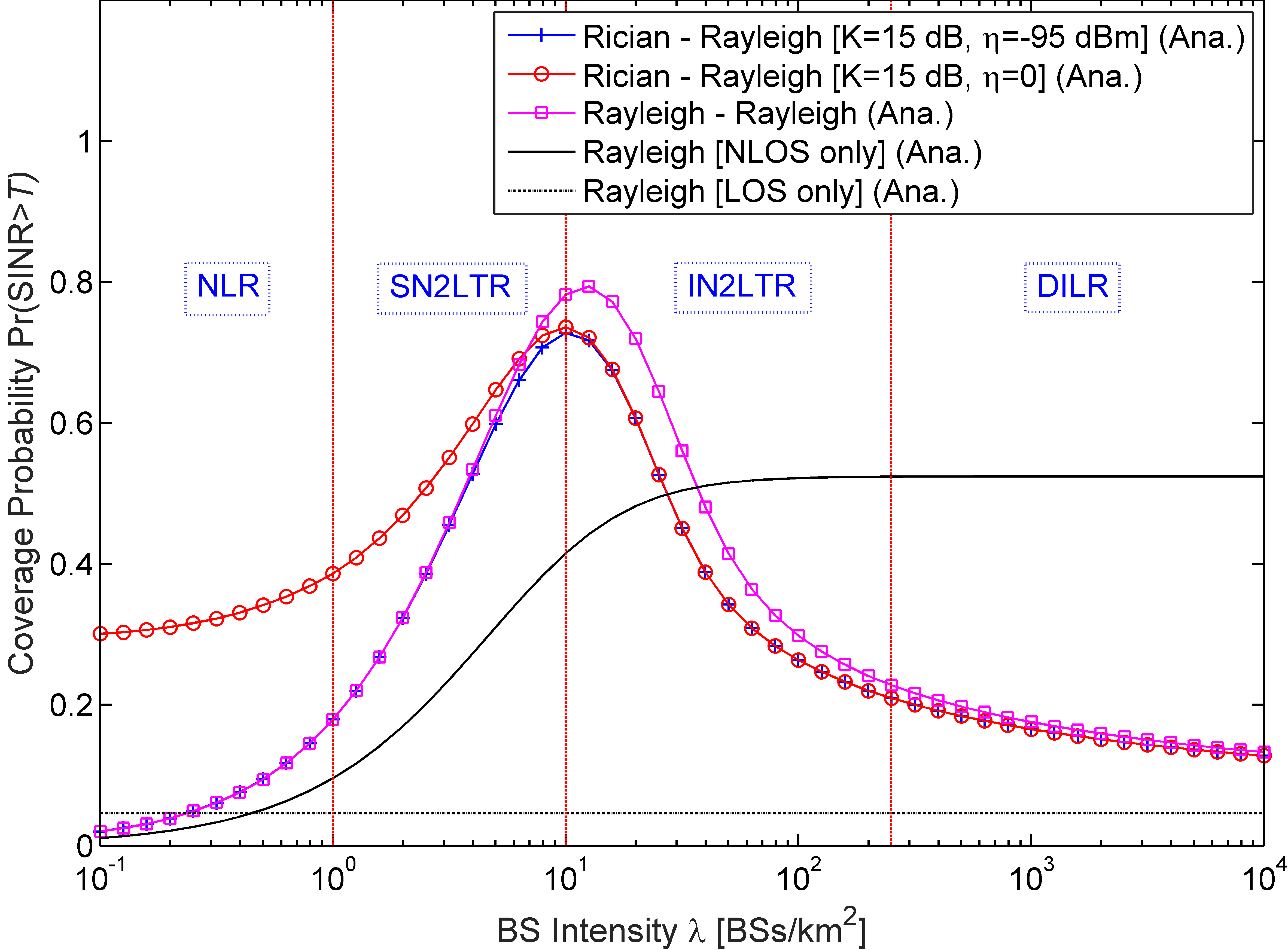}
\par\end{centering}
\caption{\label{fig:SINR-Coverage-Probability-density2}Coverage probability
vs. BS intensity $\lambda$, $T=0\textrm{ dB}.$}
\end{figure}

The results of $p_{c}\left(\lambda,T\right)$ configured with $T=0\textrm{ dB}$
are plotted in Fig. \ref{fig:SINR-Coverage-Probability-density1}
and Fig. \ref{fig:SINR-Coverage-Probability-density2}. As can be
observed from Fig. \ref{fig:SINR-Coverage-Probability-density1},
the analytical results match the simulation results well, which validate
the accuracy of our theoretical analysis. Note that in the case where
both NLOS and LOS transmissions are concatenated with Rayleigh fading,
the coverage probability is the highest among the interested cases.
By contrast, in the case where NLOS transmission is concatenated with
Rayleigh fading and LOS transmission is concatenated with Rician fading
with $K=15\textrm{ dB}$, the coverage probability is the lowest.
Meanwhile we should notice that the gap between the plotted curves
is small, which suggests that multi-path fading has a minor impact
on the coverage probability performance. In the following, we develop
a more detailed analysis on the coverage probability according to
$\lambda$, i.e.,
\begin{itemize}
\item \textbf{Noise-Limited Regime (NLR): }$\lambda\leqslant1\textrm{ BSs/k\ensuremath{m^{2}}}$.
In this regime, the typical MU is likely to have a NLOS path with
the serving BS. The network in the NLR regime is very sparse and thus
the interference can be ignored compared with the thermal noise if
we use \textbf{SINR} as our performance metric. In this case, $\textrm{SINR}=\frac{S}{\eta}$
and the coverage probability will increase with the increase of $\lambda$
as the strongest received power ($S$) grows and noise power ($\eta$)
remains the same. On the other hand, if we use \textbf{SIR} as our
performance metric, the SIR coverage probability shows a flat trail
in this regime as $\lambda$ increases (see Fig. \ref{fig:SINR-Coverage-Probability-density2}).
This is because the increase in the received signal power is almost
counterbalanced by the increase in the aggregate interference power.
Besides, as the aggregate interference power is smaller than noise
power, the SIR coverage probability is larger than the SINR coverage
probability.
\item \textbf{Signal NLOS-to-LOS-Transition Regime (SN2LTR): }$\lambda\in(1,12]\textrm{ BSs/k\ensuremath{m^{2}}}$.
In this regime, when $\lambda$ is small, the typical MU has a higher
probability to connect to a NLOS BS; while when $\lambda$ becomes
larger, the typical MU has an increasingly higher probability to connect
to a LOS BS. That is to say, with the increase of $\lambda$, the
typical MU is more likely to be associated with a LOS BS, i.e., the
received signal transforms from a NLOS path to a LOS one. Even though
the associated BS is LOS, the majority of interfering BSs are still
NLOS in this regime and thus the SINR (or SIR) coverage probability
continues growing. Besides, from this regime on, noise power has a
negligible impact on coverage performance, i.e., the SCN becomes interference-limited.
\item \textbf{Interference NLOS-to-LOS-Transition Regime (IN2LTR): }$\lambda\in(12,250]\textrm{ BSs/k\ensuremath{m^{2}}}$.
In this regime, the typical MU is connected to a LOS BS with a high
probability. However, different from the situation in the SN2LTR,
the majority of interfering BSs experience transitions from NLOS to
LOS path, which causes much more severe interference to the typical
MU compared with interfering BSs with NLOS paths. As a result, the
SINR (or SIR) coverage probability decreases with the increase of
$\lambda$ because the transition of interference from NLOS path to
LOS path causes a larger increase in interference compared with that
in signal. Note that in this regime the coverage probability performance
in our model exhibits a huge difference from that of the analysis
in \cite{Andrews11A}, which are indicated as ``NLOS only'' and
``LOS only'' in Fig. \ref{fig:SINR-Coverage-Probability-density2}.
\item \textbf{Dense Interference-Limited Regime (DILR):} $\lambda>250\textrm{ BSs/k\ensuremath{m^{2}}}$.
In this regime, the network is extremely dense and LOS BSs dominate
the SCNs. The SINR (or SIR) coverage probability becomes stable with
the increase of BS intensity as any increase in the received LOS BS
signal power is counterbalanced by the increase in the aggregate LOS
BS interference power.
\end{itemize}
\begin{rem}
Note that the boundaries between two adjacent regimes are quite qualitative
which are worth further investigation.
\end{rem}

\begin{rem}
Note that the qualitative analysis of the coverage probability performance
is in line with the findings in \cite{Ding16Performance} (see the
curve with square markers in Fig. 2). 
\end{rem}

\subsection{Discussion on the Analytical Results of $\textrm{ASE}\left(\lambda\right)$}

\begin{figure}
\begin{centering}
\includegraphics[width=8.3cm]{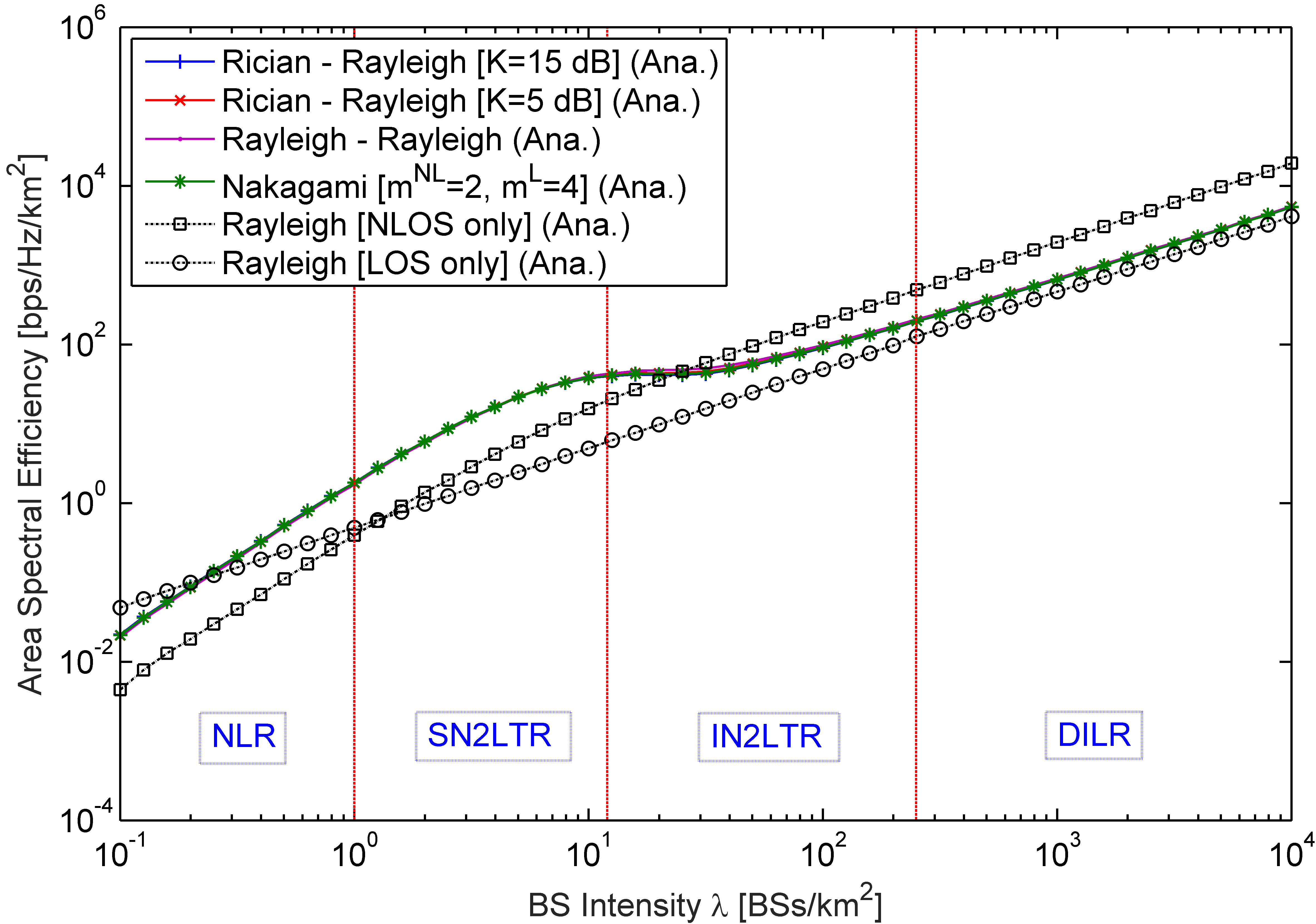}
\par\end{centering}
\caption{\label{fig:ASE}ASE vs. BS intensity $\lambda$, $\eta=-95\textrm{ dBm}$,
$T=0\textrm{ dB}.$}
\end{figure}

In this subsection, the ASE with $T=0\textrm{ dB}$ is evaluated analytically,
as $\textrm{ASE}\left(\lambda\right)$ is a function of $p_{c}\left(\lambda,T\right)$
which is analytically investigated in Eq. \eqref{eq:ASE}.

Fig. \ref{fig:ASE} illustrates the ASE with different fading models
vs. $\lambda$. It is found that the ASE of the SCN incorporating
both NLOS and LOS transmissions reveal a deviation from that of the
analysis considering NLOS (or LOS) transmissions only \cite{Andrews11A}.
Specifically, when the SCN is sparse and thus in the NLR or the SN2LTR,
the ASE quickly increases with $\lambda$ because the network is generally
noise-limited, and thus adding more small cells immensely benefits
the ASE. When the network becomes dense, i.e., $\lambda$ enters the
IN2LTR, which is the practical range of $\lambda$ for the existing
4G networks and the future 5G networks, the trend of the ASE performance
is very interesting. First, when $\lambda\in(12,50]\textrm{ BSs/k\ensuremath{m^{2}}}$,
the ASE exhibits a slowing-down in the rate of growth due to the fast
decrease of the coverage probability at $\lambda\in(12,50]\textrm{ BSs/k\ensuremath{m^{2}}}$,
as shown in Figs. \ref{fig:SINR-Coverage-Probability-density1} and
\ref{fig:SINR-Coverage-Probability-density2}. Second, when $\lambda>50\textrm{ BSs/k\ensuremath{m^{2}}}$,
the ASE will pick up the growth rate since the decrease of the coverage
probability becomes a minor factor compared with the increase of $\lambda$.
When the SCN is extremely dense, e.g., $\lambda$ is in the DILR,
the ASE exhibits a nearly linear trajectory with regard to $\lambda$
because both the signal power and the interference power are now LOS
dominated, and thus statistically stable as explained before. Moreover,
it can be observed that the change of the multi-path fading model
has a minor impact on the ASE performance compared with the change
of the path loss model.

\section{\label{sec:Conclusions-and-Future}Conclusions and Future Work}

In this paper, we proposed a unified framework to analyze the performance
of the SCNs. In our analysis, we considered a practical path loss
model that accounts for both NLOS and LOS transmissions. Furthermore,
we adopted a generalized fading model, in which Rayleigh fading, Rician
fading and Nakagami-$m$ fading can be treated in a unified framework.
Different from existing work that does not differentiate NLOS and
LOS transmissions, our results show that the co-existence of NLOS
and LOS transmissions have a significant impact on the coverage probability
and the ASE performance. Furthermore, our results establish for the
first time that the performance of the SCNs can be divided into four
regimes, according to the intensity of BSs, where in each regime the
performance is dominated by different factors. 

In our future work, both shadow fading and multi-path fading will
be considered in our analysis which is more practical for the real
network. Furthermore, heterogeneous networks (HetNets) incorporating
both NLOS and LOS transmissions will also be investigated.

\section*{Acknowledgment}

The authors would like to acknowledge the support from the NSFC Major
International Joint Research Project (Grant No. 61210002), Hubei Provincial
Science and Technology Department under Grant 2016AHB006, the Fundamental
Research Funds for the Central Universities under the grant 2015XJGH011.
This research is partially supported by the EU FP7-PEOPLE-IRSES, project
acronym CROWN (Grant No. 610524), China International Joint Research
Center of Green Communications and Networking (No. 2015B01008). The
work is also supported by the UTS-HUST Joint Research Center of Mobile
Communications (JRCMC) and the China Scholarship Council (CSC).

\section*{Appendix: Proof of Theorem \ref{thm:Pcoverage }}

If we denote $\overline{R_{i}^{\textrm{NL}}}=R_{i}\cdot\left(B^{\textrm{NL}}h^{\textrm{NL}}\right)^{-1/\alpha^{\textrm{NL}}}$
and $\overline{R_{i}^{\textrm{L}}}=R_{i}\cdot\left(B^{\textrm{L}}h^{\textrm{L}}\right)^{-1/\alpha^{\textrm{L}}}$,
$\left\{ \overline{R_{i}^{\textrm{NL}}}\right\} _{i\in\mathbb{N}}$
(or $\left\{ \overline{R_{i}^{\textrm{L}}}\right\} _{i\in\mathbb{N}}$)
remains a PPP denoted by $\overline{\Phi^{\textrm{NL}}}$ (or $\overline{\Phi^{\textrm{L}}}$
) according to the displacement theorem \cite{Baccelli09Stochastic}.
The intensity measure and intensities of $\overline{\Phi^{\textrm{NL}}}$
and $\overline{\Phi^{\textrm{L}}}$ are given by Eq. \eqref{eq:Measure_N}
- Eq. \eqref{eq:Intensity_L}. The proof can be referred to \cite{Blaszczyszyn13Using,Yang17Performance},
which is omitted here. Next, we give the main proof of Theorem \ref{thm:Pcoverage }.

By invoking the law of total probability, the coverage probability
can be divided into two parts, i.e., $p_{c}^{\textrm{NL}}\left(\lambda,T\right)$
and $p_{c}^{\textrm{L}}\left(\lambda,T\right)$, which denotes the
conditional coverage probability given that the typical MU is associated
with a BS in $\Phi^{\textrm{NL}}$ and $\Phi^{\textrm{L}}$, respectively.
Moreover, denote by $\mathcal{P}^{\textrm{NL}}$ and $\mathcal{P}^{\textrm{L}}$
the strongest received signal power from BS in $\Phi^{\textrm{NL}}$
and $\Phi^{\textrm{L}}$, i.e., $\mathcal{P}^{\textrm{NL}}=\max\left(P_{i}^{\textrm{NL}}\right)$
and $\mathcal{P}^{\textrm{L}}=\max\left(P_{i}^{\textrm{L}}\right)$,
respectively. Then by using the law of total probability, $p_{c}^{\textrm{L}}\left(\lambda,T\right)$
can be computed by
\begin{align}
p_{c}^{\textrm{L}}\left(\lambda,T\right) & =\Pr\left[\left(\textrm{SINR}^{\textrm{L}}>T\right)\cap\left(\mathcal{P}^{\textrm{L}}>\mathcal{P}^{\textrm{NL}}\right)\cap\mathcal{Y}^{\textrm{L}}\right]\nonumber \\
 & =\mathbb{E}_{\mathcal{Y}^{\textrm{L}}}\biggl\{\underset{\textrm{II}}{\underbrace{\Pr\left[\left.\textrm{SINR}^{\textrm{L}}>T\right|\left(\mathcal{P}^{\textrm{L}}>\mathcal{P}^{\textrm{NL}}\right)\cap\mathcal{Y}^{\textrm{L}}\right]}}\nonumber \\
 & \quad\,\times\underset{\textrm{I}}{\underbrace{\Pr\left[\left.\mathcal{P}^{\textrm{L}}>\mathcal{P}^{\textrm{NL}}\right|\mathcal{Y}^{\textrm{L}}\right]}}\biggr\},\label{eq:proof_pcL}
\end{align}
where $\mathcal{Y}^{\textrm{L}}$ is the equivalent distance between
the typical MU and the BS providing the strongest received signal
power to the typical MU in $\Phi^{\textrm{L}}$, i.e., $\mathcal{Y}^{\textrm{L}}=\underset{\overline{R_{i}^{\textrm{L}}}\in\overline{\Phi^{\textrm{L}}}}{\arg\max}\left(\overline{R_{i}^{\textrm{L}}}\right)^{-\alpha^{\textrm{L}}}$,
and also note that $\mathcal{P}^{\textrm{L}}=\left(\mathcal{Y}^{\textrm{L}}\right)^{-\alpha^{\textrm{L}}}$.
Besides, Part I guarantees that the typical MU is connected to a LOS
BS and Part II denotes the coverage probability conditioned on the
proposed cell association scheme in Eq. \eqref{eq:maxrecpower}. Next,
Part I and Part II will be respectively derived separately. For Part
I,
\begin{align}
 & \quad\,\Pr\left[\left.\mathcal{P}^{\textrm{L}}>\mathcal{P}^{\textrm{NL}}\right|\mathcal{Y}^{\textrm{L}}\right]\nonumber \\
 & =\Pr\left[\left.\left(\mathcal{Y}^{\textrm{L}}\right)^{-\alpha^{\textrm{L}}}>\left(\mathcal{Y}^{\textrm{NL}}\right)^{-\alpha^{\textrm{NL}}}\right|\mathcal{Y}^{\textrm{L}}\right]\nonumber \\
 & =\Pr\left[\left.\mathcal{Y}^{\textrm{NL}}>\left(\mathcal{Y}^{\textrm{L}}\right)^{\alpha^{\textrm{L}}/\alpha^{\textrm{NL}}}\right|\mathcal{Y}^{\textrm{L}}\right]\nonumber \\
 & \overset{\left(a\right)}{=}\exp\left[-\Lambda^{\textrm{NL}}\left(\left[0,\left(\mathcal{Y}^{\textrm{L}}\right)^{\alpha^{\textrm{L}}/\alpha^{\textrm{NL}}}\right]\right)\right],\label{eq:proof_PL g PN}
\end{align}
where $\mathcal{Y}^{\textrm{NL}}$, similar to the definition of $\mathcal{Y}^{\textrm{L}}$,
is the equivalent distance between the typical MU and the BS providing
the strongest received signal power to the typical MU in $\Phi^{\textrm{NL}}$,
i.e., $\mathcal{Y}^{\textrm{NL}}=\underset{\overline{R_{i}^{\textrm{NL}}}\in\overline{\Phi^{\textrm{NL}}}}{\arg\max}\left(\overline{R_{i}^{\textrm{NL}}}\right)^{-\alpha^{\textrm{NL}}}$,
and also note that $\mathcal{P}^{\textrm{NL}}=\left(\mathcal{Y}^{\textrm{NL}}\right)^{-\alpha^{\textrm{NL}}}$,
and $\left(a\right)$ follows from the void probability of a PPP.

For Part II, we know that $\textrm{SINR}=\frac{\mathcal{P}}{I+\eta}=\frac{\mathcal{P}}{I^{\textrm{NL}}+I^{\textrm{L}}+\eta},$
where $I^{\textrm{NL}}$ and $I^{\textrm{L}}$ denote the aggregate
interference from NLOS BSs and LOS BSs, respectively. The conditional
coverage probability is derived as follows
\begin{align}
 & \quad\,\Pr\left[\left.\textrm{SINR}^{\textrm{L}}>T\right|\left(\mathcal{P}^{\textrm{L}}>\mathcal{P}^{\textrm{NL}}\right)\cap\mathcal{Y}^{\textrm{L}}\right]\nonumber \\
 & =\Pr\left[\left.\frac{1}{\textrm{SINR}^{\textrm{L}}}<\frac{1}{T}\right|\left(\mathcal{P}^{\textrm{L}}>\mathcal{P}^{\textrm{NL}}\right)\cap\mathcal{Y}^{\textrm{L}}\right]\nonumber \\
 & \overset{\left(a\right)}{=}\int_{x=0}^{1/T}\int_{\omega=-\infty}^{\infty}\frac{e^{-j\omega x}}{2\pi}\mathcal{F}_{\frac{1}{\textrm{SINR}^{\textrm{L}}}}\left(\omega\right)\textrm{d}\omega\textrm{d}x\nonumber \\
 & =\int_{\omega=-\infty}^{\infty}\left[\frac{1-e^{-j\omega/T}}{2\pi j\omega}\right]\mathcal{F}_{\frac{1}{\textrm{SINR}^{\textrm{L}}}}\left(\omega\right)\textrm{d}\omega,\label{eq:proof_SINR}
\end{align}
where $\textrm{SINR}^{\textrm{L}}$ denotes the SINR when the typical
MU is associated with a LOS BS, the inner integral in $\left(a\right)$
is the conditional PDF of $\frac{1}{\textrm{SINR}^{\textrm{L}}}$,
and $\mathcal{F}_{\frac{1}{\textrm{SINR}^{\textrm{L}}}}\left(\omega\right)$
denotes the conditional characteristic function of $\frac{1}{\textrm{SINR}^{\textrm{L}}}$
which is given by
\begin{align}
 & \mathcal{F}_{\frac{1}{\textrm{SINR}^{\textrm{L}}}}\left(\omega\right)=\mathbb{E}_{\Phi}\left[\left.\exp\left(j\omega\frac{1}{\textrm{SINR}^{\textrm{L}}}\right)\right|\left(\mathcal{P}^{\textrm{L}}>\mathcal{P}^{\textrm{NL}}\right)\cap\mathcal{Y}^{\textrm{L}}\right]\nonumber \\
 & =\mathbb{E}_{\Phi}\left[\left.e^{j\omega\left(I^{\textrm{NL}}+I^{\textrm{L}}+\eta\right)\left(\mathcal{Y}^{\textrm{L}}\right)^{\alpha^{\textrm{L}}}}\right|\left(\mathcal{P}^{\textrm{L}}>\mathcal{P}^{\textrm{NL}}\right)\cap\mathcal{Y}^{\textrm{L}}\right]\nonumber \\
 & \overset{\left(a\right)}{=}\mathbb{E}_{\Phi^{\textrm{NL}}}\left\{ \left.\exp\left[j\omega I^{\textrm{NL}}\cdot\left(\mathcal{Y}^{\textrm{L}}\right)^{\alpha^{\textrm{L}}}\right]\right|\left(\mathcal{P}^{\textrm{L}}>\mathcal{P}^{\textrm{NL}}\right)\cap\mathcal{Y}^{\textrm{L}}\right\} \nonumber \\
 & \quad\,\times\mathbb{E}_{\Phi^{\textrm{L}}}\left\{ \left.\exp\left[j\omega I^{\textrm{L}}\cdot\left(\mathcal{Y}^{\textrm{L}}\right)^{\alpha^{\textrm{L}}}\right]\right|\left(\mathcal{P}^{\textrm{L}}>\mathcal{P}^{\textrm{NL}}\right)\cap\mathcal{Y}^{\textrm{L}}\right\} \nonumber \\
 & \quad\,\times e^{j\omega\eta\left(\mathcal{Y}^{\textrm{L}}\right)^{\alpha^{\textrm{L}}}},\label{eq:Characteristic_1/SINR}
\end{align}
where $\left(a\right)$ comes from the facts that $\Phi=\Phi^{\textrm{NL}}\cup\Phi^{\textrm{L}}$
and the mutual independence of $\Phi^{\textrm{NL}}$ and $\Phi^{\textrm{L}}$.
Now by applying concepts from stochastic geometry, we will derive
the term $\mathbb{E}_{\Phi^{\textrm{NL}}}\left\{ \left.\exp\left[j\omega I^{\textrm{NL}}\left(\mathcal{Y}^{\textrm{L}}\right)^{\alpha^{\textrm{L}}}\right]\right|\left(\mathcal{P}^{\textrm{L}}>\mathcal{P}^{\textrm{NL}}\right)\cap\mathcal{Y}^{\textrm{L}}\right\} $
in Eq. \eqref{eq:Characteristic_1/SINR} as follows
\begin{align}
 & \quad\,\mathbb{E}_{\Phi^{\textrm{NL}}}\left\{ \left.\exp\left[j\omega I^{\textrm{NL}}\cdot\left(\mathcal{Y}^{\textrm{L}}\right)^{\alpha^{\textrm{L}}}\right]\right|\left(\mathcal{P}^{\textrm{L}}>\mathcal{P}^{\textrm{NL}}\right)\cap\mathcal{Y}^{\textrm{L}}\right\} \nonumber \\
 & \overset{\left(a\right)}{=}\mathbb{E}_{\overline{\Phi^{\textrm{NL}}}}\Biggl\{\left.\underset{i:\overline{R_{i}^{\textrm{NL}}}\in\overline{\Phi^{\textrm{NL}}}'}{\prod}\exp\left[j\omega\cdot\left(\mathcal{Y}^{\textrm{L}}\right)^{\alpha^{\textrm{L}}}\left(\overline{R_{i}^{\textrm{NL}}}\right)^{-\alpha^{\textrm{NL}}}\right]\right|\bigl(\mathcal{P}^{\textrm{L}}\nonumber \\
 & \quad\,>\mathcal{P}^{\textrm{NL}}\bigr)\cap\mathcal{Y}^{\textrm{L}}\Biggr\}\nonumber \\
 & \overset{\left(b\right)}{=}\exp\left\{ \int_{t=\left(\mathcal{Y}^{\textrm{L}}\right)^{\alpha^{\textrm{L}}/\alpha^{\textrm{NL}}}}^{\infty}\left[e^{j\omega\left(\mathcal{Y}^{\textrm{L}}\right)^{\alpha^{\textrm{L}}}t^{-\alpha^{\textrm{NL}}}}-1\right]\lambda^{\textrm{NL}}\left(t\right)\textrm{d}t\right\} ,\label{eq:proof_E_N}
\end{align}
where in $\left(a\right)$, $\overline{\Phi^{\textrm{NL}}}'=\overline{\Phi^{\textrm{NL}}}\setminus b\left(0,\left(\mathcal{Y}^{\textrm{L}}\right)^{\alpha^{\textrm{L}}/\alpha^{\textrm{NL}}}\right)$
and $\overline{R_{i}^{\textrm{NL}}}\in\overline{\Phi^{\textrm{NL}}}'$
can guarantee the condition that $\mathcal{P}^{\textrm{L}}>\mathcal{P}^{\textrm{NL}}$,
and $\left(b\right)$ is obtained by applying the probability generating
functional (PGFL) \cite[Eq. (3)]{Andrews11A} of the PPP. Similarly,
the term $\mathbb{E}_{\Phi^{\textrm{L}}}\left\{ \left.\exp\left[j\omega I^{\textrm{L}}\cdot\left(\mathcal{Y}^{\textrm{L}}\right)^{\alpha^{\textrm{L}}}\right]\right|\left(\mathcal{P}^{\textrm{L}}>\mathcal{P}^{\textrm{NL}}\right)\cap\mathcal{Y}^{\textrm{L}}\right\} $
in Eq. \eqref{eq:Characteristic_1/SINR} is given by
\begin{align}
 & \quad\,\mathbb{E}_{\Phi^{\textrm{L}}}\left\{ \left.\exp\left[j\omega I^{\textrm{L}}\cdot\left(\mathcal{Y}^{\textrm{L}}\right)^{\alpha^{\textrm{L}}}\right]\right|\left(\mathcal{P}^{\textrm{L}}>\mathcal{P}^{\textrm{NL}}\right)\cap\mathcal{Y}^{\textrm{L}}\right\} \nonumber \\
 & =\exp\left\{ \int_{t=\mathcal{Y}^{\textrm{L}}}^{\infty}\left[e^{j\omega\left(\mathcal{Y}^{\textrm{L}}/t\right)^{\alpha^{\textrm{L}}}}-1\right]\lambda^{\textrm{L}}\left(t\right)\textrm{d}t\right\} \text{.}\label{eq:proof_E_L}
\end{align}
Then the product of Part I and Part II in Eq. \eqref{eq:proof_pcL}
can be obtained by substituting them with Eq. \eqref{eq:proof_PL g PN}
\textendash{} Eq. \eqref{eq:proof_E_L}.

Finally, note that the value of $p_{c}^{\textrm{L}}\left(\lambda,T\right)$
in Eq. \eqref{eq:proof_pcL} should be calculated by taking the expectation
with respect to $\mathcal{Y}^{\textrm{L}}$ in terms of its PDF, which
is given as follows
\begin{equation}
f_{\mathcal{Y}^{\textrm{L}}}\left(y\right)=\frac{\textrm{d}}{\textrm{d}y}\left[1-\Pr\left(\mathcal{Y}^{\textrm{L}}>y\right)\right]=\lambda^{\textrm{L}}\left(y\right)\exp\left[-\Lambda^{\textrm{L}}\left(\left[0,y\right]\right)\right].
\end{equation}

Given that the typical MU is connected to a NLOS BS, the conditional
coverage probability $p_{c}^{\textrm{NL}}\left(\lambda,T\right)$
can be derived using the similar way as the above. In this way, the
coverage probability is obtained by $p_{c}\left(\lambda,T\right)=p_{c}^{\textrm{L}}\left(\lambda,T\right)+p_{c}^{\textrm{NL}}\left(\lambda,T\right)$.
Thus the proof is completed.

\bibliographystyle{IEEEtran}
\bibliography{BY}

\end{document}